\documentclass[runningheads]{llncs}
\usepackage[T1]{fontenc}
\usepackage{graphicx}
\usepackage{array}
\usepackage{booktabs}
\usepackage{multirow}
\usepackage{amsmath}
\usepackage{amssymb}
\usepackage{tikz}
\usepackage{ulem}
\usetikzlibrary{shapes.geometric, arrows, positioning}
\usepackage{soul, color, xcolor}
\sethlcolor{yellow}
\newcommand{\SystemName}{\texttt{GESR}}
\begin{document}
\raggedbottom

\title{GESR: Graph-Based Edge Semantic Reconstruction for Stealthy Communication Detection with Benign-Only Training}
\titlerunning{\SystemName{} for Stealthy Communication Detection}
\author{
Henghui Xu\inst{1,2} \and
Yuchen Zhang\inst{1,2} \and
Xiaobo Ma\inst{1,2,3}
}

\authorrunning{H. Xu et al.}

\institute{
MOE Key Laboratory for Intelligent Networks and Network Security, Xi'an Jiaotong University, Xi'an, China
\and
Faculty of Electronic and Information Engineering, Xi'an Jiaotong University, Xi'an, China
\and
Shaanxi Province Key Laboratory of Computer Network, Xi'an Jiaotong University, Xi'an, China\\
\email{
henghuixu.academic@gmail.com,
yvchen.zhang@outlook.com,
xma.cs@xjtu.edu.cn
}
}
\maketitle              

\begin{abstract}
Detecting stealthy malicious communications from flow logs under benign-only training remains a critical challenge in network security.
Malicious communications often camouflage as normal traffic like standard HTTPS flows.
Conventional intrusion detectors rely strictly on known labeled attacks.
Alternatively, they score flows completely independently.
These approaches fail against sparse and context-dependent suspicious activity.
To capture this essential context, graph anomaly detectors have been introduced to add valuable relational information to the analysis.
However, existing methods fail to test the structural consistency of specific communication edges.
To overcome these fundamental limitations, we present \SystemName{}, a novel graph-based framework for detecting suspicious communications and anomalous hosts under a benign-only training setting.
\SystemName{} models complex network activity as attributed communication graphs.
It cleverly reconstructs edge semantics entirely from local structural context rather than isolated features.
This non-intuitive design forces the framework to predict expected communication patterns from neighborhood topologies.
Attackers cannot easily manipulate this deep structural dependency.
The model then converts the resulting structural inconsistencies into host-level anomaly scores.
It utilizes robust Median Absolute Deviation (MAD) calibration for this final step.
We evaluate \SystemName{} extensively on CTU-13 and CICIDS2017 datasets.
These evaluations strictly impose tight false-positive operating constraints.
On CICIDS2017, \SystemName{} achieves an outstanding ROC-AUC of 0.9753.
It also yields a high TPR of 0.8569 at a strict 5\% FPR threshold.
\SystemName{} consistently outperforms existing methods across both evaluated benchmarks.
The results prove that structure-conditioned edge reconstruction is a credible direction for practical intrusion detection.

\keywords{intrusion detection \and graph anomaly detection \and benign-only detection \and self-supervised learning}
\end{abstract}

\section{Introduction}

Command-and-control (C2) communication remains a critical threat in network security. 
Adversaries use these hidden channels to maintain control over compromised hosts and coordinate subsequent malicious activities. 
Disrupting this control channel effectively limits an attacker's capabilities. 
Modern C2 traffic increasingly employs protocol mimicry, encryption, and low-volume communication patterns to camouflage as normal network activity~\cite{ref_c2_evasion}. 
Identifying these stealthy communications among massive benign flows is extremely difficult. 
The core obstacle lies in the severe statistical overlap between benign and malicious traffic when viewed as isolated flows.
Their abnormality emerges only after researchers place the flows in the broader context of network interactions. 
Various detection methods attempt to address this challenge. However, they have inherent limitations in identifying highly camouflaged threats.

\noindent\textbf{Neglect of Relational Context in Flow-based Analysis.} 
Flow-based and feature-based methods attempt to model statistical properties of network traffic~\cite{ref_anomaly_survey,ref_kitsune}.
These approaches treat each flow as an independent observation. 
Consequently, they completely neglect the relational context among communicating entities. 
Without such context, stealthy C2 communications can easily mimic standard HTTPS flows and evade detection.
They look completely benign when viewed as isolated events. 
A parallel line of encrypted-traffic studies uses TLS metadata and contextual flow information to identify malicious sessions even when decryption is unavailable~\cite{ref_tls_context,ref_tls_nonstationary,ref_tls_usage}.
These approaches can capture certain traffic patterns. 
However, they primarily rely on flow-level observations and local statistical features. 
They fail to explicitly model correlations among communicating entities~\cite{ref_maldiscovery,ref_stgraph}.
Consequently, suspicious communications that closely resemble benign traffic may evade detection under such representations.

\noindent\textbf{Coarse-grained Modeling in Graph-based Detection.} 
Graph-based methods model network interactions as graphs to incorporate relational structure~\cite{ref_graph_based}.
Many existing approaches focus heavily on node-level representations or treat the graph as a whole. 
They lack fine-grained modeling of individual communication edges. 
A compromised host might exhibit normal overall traffic volumes but establish a structural anomaly with a specific C2 server. 
Therefore, relying solely on node-level or global structural representations is insufficient to capture anomalies arising from structural inconsistency.

To overcome the limitations above, a robust detection framework must achieve two objectives. 
First, it must capture the detailed relational context of individual communications. 
Second, it must explicitly verify the structural consistency of each connection. 
Fulfilling such two objectives is not easy. 
Stealthy traffic patterns easily blend into complex background noise.
Therefore, we propose \SystemName{} (Graph-Based Edge Semantic Reconstruction). 
\SystemName{} represents network activity as communication graphs. 
It uses an edge-aware encoder to reconstruct the expected semantics of each edge entirely from its local context. 
We chose edge-level reconstruction over direct node classification because stealthy behaviors typically manifest as subtle structural anomalies between specific host pairs, rather than massive node-level volume deviations.
Finally, it scores edges by the mismatch between observed behavior and context-conditioned reconstruction. 
This design forces the model to learn expected communication patterns. 
It effectively exposes suspicious interactions missed by isolated flow statistics.

To our best knowledge, we are the first to formulate benign-only detection of suspicious communications through structure-conditioned edge semantic reconstruction. We will release our source code to facilitate future research. Our contributions are summarized as follows:

\begin{itemize}
    \item We formulate benign-only detection of suspicious communications under strict false-positive operating constraints. We use stealthy C2 traffic as the motivating threat model.
    \item We propose \SystemName{}, a novel edge-semantic graph detection framework. It reconstructs expected communication behavior from structural context and converts reconstruction inconsistencies into stable host-level risk scores.
    \item We extensively evaluate \SystemName{} on CTU-13 and CICIDS2017 datasets. The results demonstrate strong host-ranking performance in the benign-only setting.
\end{itemize}

\begin{figure}[t]
\centering
\includegraphics[width=\textwidth]{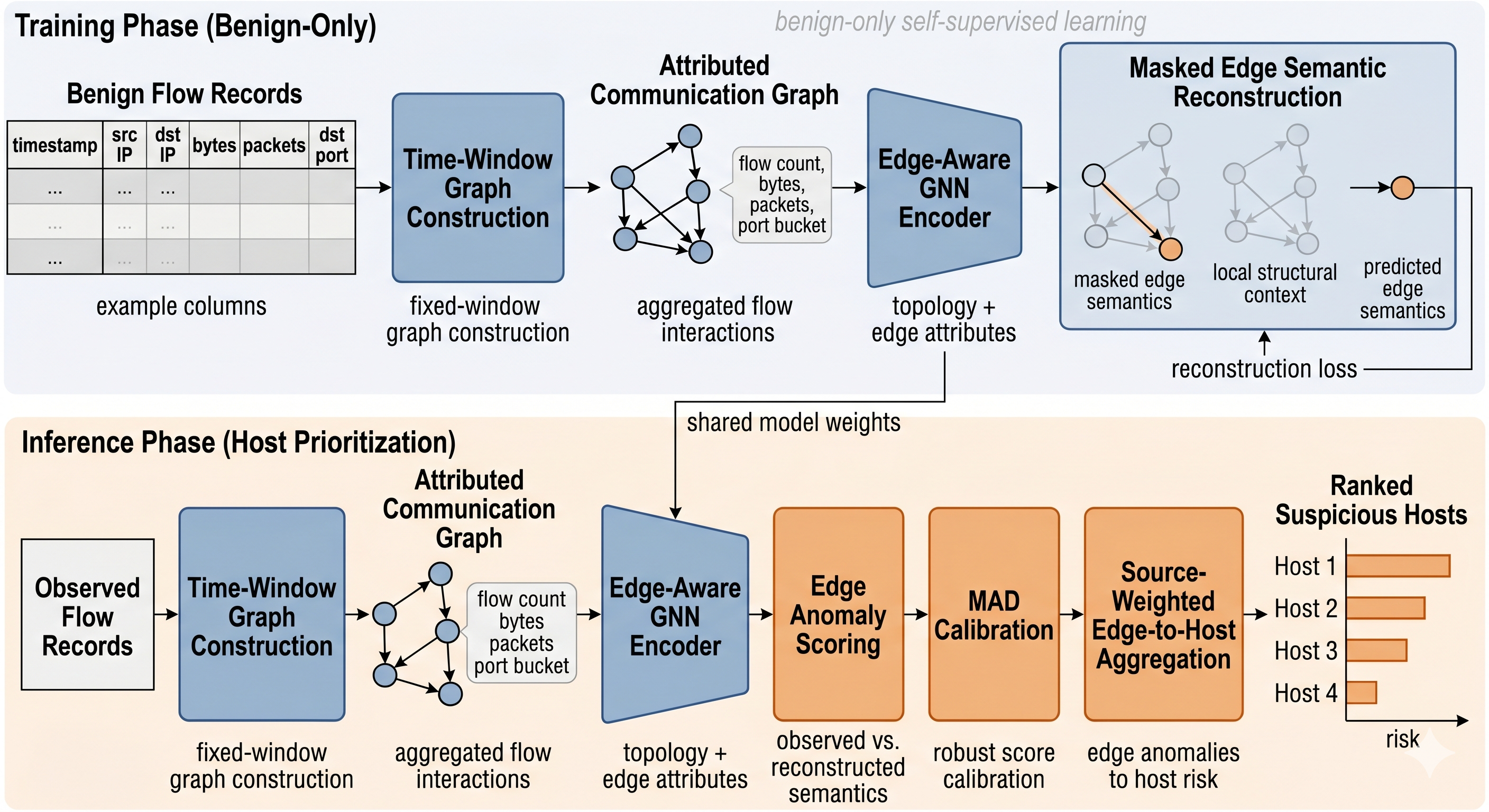}
\caption{\SystemName{} system overview. During training, \SystemName{} builds attributed communication graphs from benign flow records and learns structure-conditioned edge semantic reconstruction with an edge-aware encoder. During inference, the same encoder scores anomalous communications, calibrates edge scores, and aggregates them to produce ranked suspicious hosts.}

\label{fig:system_overview}
\end{figure}

We design \SystemName{} for offline security analytics pipelines. These pipelines ingest flow-level records rather than raw packets. This allows deployment where deep packet inspection is infeasible. We assume adversaries use evasion techniques to bypass volumetric detectors. We do not model fully adaptive attackers that can arbitrarily manipulate global graph structure. Figure~\ref{fig:system_overview} summarizes the end-to-end pipeline from flow records to prioritized host alerts.

The remainder of this paper is organized as follows. Section 2 introduces the background and problem setting. Section 3 presents graph construction and anomaly scoring in \SystemName{}. Section 4 describes the experimental setup and results. Section 5 discusses implications and limitations. Section 6 reviews related work. Section 7 concludes.

\section{Background}
\subsection{Benign-Only Detection and Operational Evaluation}
Network traffic monitoring is the cornerstone of modern enterprise security defense.
However, in operational intrusion detection, labeled attacks are scarce, delayed, and often incomplete, which makes closed-world supervised training brittle once attack behavior shifts or new threats appear~\cite{ref_supervised_limits}. 
Anomaly-based intrusion detection addresses this limitation by learning a model of regular behavior and flagging deviations at inference time rather than optimizing over a fixed taxonomy of known attacks~\cite{ref_anomaly_survey,ref_ids_survey,ref_kitsune}. In this setting, the practical goal is rarely to maximize average discrimination across all thresholds. Analysts instead need a detector that can prioritize a small number of suspicious hosts without overwhelming them with false alerts.

This deployment perspective matters because intrusion datasets are typically highly imbalanced: benign entities dominate, while malicious activity is sparse. Under such class skew, ROC-AUC remains useful but can hide large differences in alert quality at the operating points that matter most in practice. Precision-recall analysis and true positive rate measured at explicitly constrained false positive rates therefore provide a more faithful view of whether a detector is operationally usable under limited alert budgets~\cite{ref_pr_curve}. We adopt this view throughout the paper and treat host ranking under strict false-positive constraints as the primary evaluation setting for \SystemName{}.

\subsection{Communication Graphs from Flow Logs}
When packet payloads are unavailable because of encryption, privacy constraints, or collection policy, flow logs remain one of the few scalable sources of network telemetry~\cite{ref_dataset_survey}. Modern C2 channels deliberately exploit this setting by using protocol mimicry, encryption, and low-volume beaconing to blend into benign traffic~\cite{ref_c2_evasion}. A common response is to represent traffic within each time window as a communication graph, where hosts become nodes and aggregated interactions become attributed edges. Graph-based anomaly detection is well suited to this setting because it models dependencies among entities rather than treating each traffic record as an independent point~\cite{ref_graph_based}.

In security settings, this general idea has been instantiated with spatial-temporal traffic graphs for encrypted malware detection~\cite{ref_stgraph}, self-supervised edge-aware flow representations~\cite{ref_anomale}, and streaming temporal graph encoders for evolving behavior~\cite{ref_edgetorrent}. Taken together, these studies show that relational structure can carry detection signal that is difficult to recover from isolated flow statistics alone.

\subsection{Edge Semantics and Host-Level Prioritization}
For stealthy communication patterns, the anomaly often lies less in the existence of a connection than in whether the observed behavior on that connection matches its structural role. We use the term edge semantics to denote the aggregated communication attributes attached to a host pair within a time window, such as traffic volume, packet statistics, and coarse service indicators. This edge-centric view differs from methods that score hosts directly from node embeddings or summarize entire windows into graph-level sketches. It becomes especially useful when individual flows appear benign in isolation but become suspicious once placed in the context of surrounding communications.

In practice, however, defenders do not triage communication edges one by one; they triage hosts. A benign-only detector therefore needs not only to score edge-level inconsistencies, but also to convert them into stable host-level rankings under limited false-positive budgets. This requirement motivates the calibration and aggregation stages in \SystemName{}: they bridge the gap between structure-conditioned edge reconstruction and the operational output that analysts ultimately consume.

\section{Method}
\subsection{Overview}
We present \SystemName{} to address the limitations of existing C2 detection methods. \SystemName{} is a graph-based anomaly detection framework that evaluates whether an observed communication edge is semantically consistent with its structural context. Instead of directly classifying traffic, \SystemName{} learns the expected behavior of each communication edge from surrounding graph structure and identifies anomalies as deviations from these structure-conditioned expectations.

Given network flows, \SystemName{} first constructs a sequence of communication graphs, where hosts are represented as nodes and aggregated communications are represented as attributed edges. Each graph is then processed by an edge-aware message-passing encoder based on GINEConv, which learns context-rich node representations from both graph topology and edge attributes. Based on the learned node states, \SystemName{} builds structure-aware edge representations and uses an edge decoder to reconstruct communication semantics, including both continuous traffic attributes and discrete port-related semantics. This design enables \SystemName{} to model communication behavior at the edge level, rather than relying solely on node- or graph-level representations.

\SystemName{} is trained in a benign-only manner through masked edge semantic reconstruction. During training, a subset of edge attributes is masked, and the model learns to recover them from graph context, thereby capturing the regularities of benign communication behavior. During inference, \SystemName{} assigns anomaly scores according to the discrepancy between observed edge semantics and their reconstructed values. These edge-level scores are then aggregated to identify suspicious hosts. In this way, \SystemName{} can surface suspicious communications that appear benign at the flow level but remain inconsistent within the broader communication structure.
\subsection{Graph Construction}
Rather than operating on raw packets, \SystemName{} ingests flow-level network traffic records. 
The datasets in this work originate from different sources and use heterogeneous formats. 
Therefore, we first normalize their schema. 
We align the key columns required for graph construction, including timestamps, IP addresses, and flow-level statistics.
Invalid entries, missing values, and non-finite values are removed or replaced during preprocessing to ensure that the resulting graph representation is numerically stable.

After schema normalization, traffic is partitioned into fixed time windows to construct a sequence of communication graphs. For each window, we generate one graph snapshot that summarizes the communication behavior observed during that period. This representation preserves short-term interaction patterns while avoiding the noise and sparsity that arise when operating directly at the packet level. The window size is set to 30 seconds for CTU-13 and 60 seconds for CICIDS2017, chosen to balance temporal locality and graph density.

For each time window, we define a directed communication graph $G_t = (V_t, E_t)$, where each node in $V_t$ represents a host identified by its IP address, and each directed edge in $E_t$ represents the aggregated communication from a source host to a destination host within that window. Multiple flows between the same ordered pair of hosts are merged into a single edge, and their statistics are aggregated. This aggregation allows \SystemName{} to model communication behavior at the interaction level rather than treating each flow independently.

Each edge is associated with a semantic feature vector that describes the observed communication behavior between the corresponding pair of hosts. In the feature configuration used for the reported \SystemName{} results, the edge representation contains nine dimensions: $\log(1+x)$ communication count, bytes, and packets; a one-hot encoding of the dominant destination-port bucket; and three derived intensity summaries (bytes per flow, packets per flow, and bytes per packet). We group destination ports into three buckets: web (80/443), DNS (53), and other. This edge-centric design is critical for \SystemName{}, since the detection target in our setting is not merely whether two hosts are connected, but whether the observed communication semantics are consistent with the structural context in which the edge appears.

Node features, conversely, summarize host-level activity. Under the same configuration, each host is represented by 51 dimensions. The baseline 32 dimensions encode outgoing and incoming aggregate statistics, combined traffic totals, and degree-normalized averages. We then append 19 shared semantic enhancements that describe neighborhood diversity, asymmetry between outgoing and incoming traffic, neighbor-normalized traffic intensity, byte-per-packet summaries, and coarse outgoing port-bucket usage. To compress heavy-tailed distributions, we apply a $\log(1+x)$ transformation to count and volume features. Together, the node features supply the context necessary for message passing, whereas the edge attributes carry the explicit communication semantics targeted for reconstruction. Table~\ref{tab:feature_profile} summarizes the feature groups used in this configuration.

\begin{table*}[t]
\centering
\caption{Feature groups in the reported \SystemName{} feature configuration. The first 32 node dimensions and first 6 edge dimensions form the baseline graph representation; the remaining dimensions are shared semantic enhancements used for both CTU-13 and CICIDS2017.}
\label{tab:feature_profile}
\small
\setlength{\tabcolsep}{3pt}
\begin{tabular}{@{}lc>{\raggedright\arraybackslash}p{8.35cm}@{}}
\toprule
Component & Dims & Contents \\
\midrule
Node (baseline) & 32 & Outgoing and incoming traffic summaries: flow count, bytes, packets, duration, rate statistics, combined totals, and degree-normalized averages. \\
Node (enhanced) & 19 & Neighborhood and asymmetry summaries: unique out/in/total neighbors, flows per neighbor, outgoing/incoming ratios for bytes, packets, and flows, neighbor-normalized intensity, byte-per-packet summaries, and outgoing port-bucket usage. \\
Edge (baseline) & 6 & $\log(1+x)$ communication count, bytes, and packets, plus a one-hot dominant destination-port bucket. \\
Edge (enhanced) & 3 & $\log(1+x)$ bytes per flow, packets per flow, and bytes per packet. \\
Targets & 3 + 1 & Continuous reconstruction of communication count, bytes, and packets plus three-class destination-port prediction (web, DNS, other). \\
\bottomrule
\end{tabular}
\end{table*}

Finally, we induce labels to support our one-class evaluation protocol. Graph-level anomaly labels denote the presence of any malicious activity within the temporal window. Node-level labels follow a source-centric attribution: a malicious flow from $v_i$ to $v_j$ marks only the initiating host $v_i$ as anomalous for that snapshot. This design reflects the practical assumption that compromised hosts actively initiate C2 communications, and thus anomalies are more appropriately attributed to the source rather than the destination.

Overall, this preprocessing pipeline converts heterogeneous flow records into a sequence of attributed communication graphs, providing \SystemName{} with both structural context and edge-level semantics for benign-only anomaly modeling.

\subsection{\SystemName{} Model}

Let $G_t = (V_t, E_t, X_t, \mathbf{E}_t)$ denote the attributed communication graph constructed from time window $t$, where $V_t$ is the host set, $E_t$ is the directed edge set, $X_t \in \mathbb{R}^{|V_t| \times d_v}$ is the node feature matrix, and $\mathbf{E}_t \in \mathbb{R}^{|E_t| \times d_e}$ is the edge attribute matrix. Each directed edge $e_{ij} \in E_t$ corresponds to the aggregated communication from host $v_i$ to host $v_j$ within the current window. Our goal is to learn a function that assigns an anomaly score to each communication edge by evaluating its consistency with the surrounding graph structure without using labeled attacks during training.

\SystemName{} models anomalous communication as \textbf{structural-semantic inconsistency}: under benign conditions, the semantic behavior of an edge should be predictable from the surrounding communication structure, whereas malicious edges are more likely to deviate from such context-conditioned expectations. Formally, \SystemName{} consists of three components: (i) a graph encoder that learns context-aware node representations, (ii) an edge semantic decoder that reconstructs expected communication behavior, and (iii) an anomaly scoring module that quantifies reconstruction inconsistency.

\subsubsection{Structure-Aware Node Encoding}

We first encode each communication graph with an edge-aware graph neural network. Since edge semantics are essential in our setting, the encoder must incorporate both topology and edge attributes during message passing. We therefore adopt a two-layer \textbf{GINEConv} backbone~\cite{ref_gine}, which extends edge-aware message passing by injecting edge features into neighborhood aggregation.

Given the input node feature $x_i$ for node $v_i$, we first project it into the hidden space:
\begin{equation}
h_i^{(0)} = \phi_0(x_i),
\end{equation}
where $\phi_0(\cdot)$ is a learnable linear projection followed by normalization and nonlinearity.

For the $l$-th GINE layer, node representations are updated as
\begin{equation}
h_i^{(l)} = \mathrm{GINE}^{(l)}\!\left(
h_i^{(l-1)},
\left\{ h_j^{(l-1)}, e_{ji} \mid (v_j, v_i) \in E_t \right\}
\right),
\end{equation}
where $e_{ji}$ denotes the edge attribute associated with the incoming communication from $v_j$ to $v_i$. Intuitively, the encoder learns a representation for each host by aggregating its neighboring communications and their semantics, allowing node embeddings to reflect both local interaction patterns and communication context.

To preserve information across different receptive fields, we fuse multi-level representations rather than using only the final layer output. Specifically, for each node we combine the projected input representation and the intermediate GNN outputs:
\begin{equation}
z_i = \phi_f\!\left(\left[ h_i^{(0)} \,\|\, h_i^{(1)} \,\|\, h_i^{(2)} \right]\right),
\end{equation}
where $\left[\cdot \,\|\, \cdot\right]$ denotes concatenation and $\phi_f(\cdot)$ is a learnable fusion MLP. The resulting $z_i$ serves as the final structure-aware representation of node $v_i$. These node representations are not used directly for node-level prediction; instead, they provide contextual signals for modeling edge-level communication behavior in subsequent stages.

In addition to node-level context, \SystemName{} also computes a graph-level summary for each window:
\begin{equation}
g_t = \frac{1}{|V_t|} \sum_{v_i \in V_t} z_i,
\end{equation}
which encodes the overall communication state of the current graph and provides a coarse contextual prior for edge modeling.

\subsubsection{Structure-Aware Edge Representation}

The key object in \SystemName{} is not the node but the communication edge. For each directed edge $e_{ij}$, we construct an edge representation that jointly captures the source and destination host states, the observed edge attributes, and the structural role of the edge in the surrounding graph.

Formally, let $\tilde{e}_{ij}$ denote the possibly masked input edge attributes used during training. We define the edge representation as
\begin{equation}
r_{ij} =
\left[
z_i \,\|\, z_j \,\|\, \tilde{e}_{ij} \,\|\, |z_i - z_j| \,\|\, (z_i \odot z_j) \,\|\, g_t
\right],
\end{equation}

where $|z_i - z_j|$ captures the discrepancy between the two endpoints, $z_i \odot z_j$ denotes element-wise product, and $g_t$ provides graph-level context. This formulation allows \SystemName{} to relate the observed edge semantics to both endpoint roles and their surrounding structural context, making deviations from such relationships directly measurable.

This construction is motivated by two considerations. First, whether an edge is anomalous depends not only on its observed attributes, but also on the identities and roles of its two endpoints. Second, subtle C2 edges may appear locally benign, yet become suspicious when compared against the structural expectations implied by nearby nodes and the global communication pattern. The edge representation $r_{ij}$ is designed to expose exactly such inconsistencies.

\subsubsection{Edge Semantic Reconstruction}

Instead of directly learning a binary classifier, \SystemName{} learns to reconstruct the expected semantics of each communication edge from its structure-aware representation. This design aligns with the benign-only training setting: the model is optimized to explain normal communications, and edges that cannot be well explained are treated as suspicious.

We decompose edge semantics into two parts:

\begin{itemize}
    \item \textbf{continuous semantics}, corresponding to quantitative traffic behavior such as communication count, bytes, and packets;
    \item \textbf{discrete semantics}, corresponding to coarse-grained port-related categories.
\end{itemize}

Accordingly, the decoder contains two prediction heads. Given $r_{ij}$, the continuous reconstruction head predicts
\begin{equation}
\hat{y}^{\mathrm{reg}}_{ij} = \psi_{\mathrm{reg}}(r_{ij}),
\end{equation}
and the discrete classification head predicts
\begin{equation}
\hat{y}^{\mathrm{cls}}_{ij} = \psi_{\mathrm{cls}}(r_{ij}),
\end{equation}
where $\psi_{\mathrm{reg}}(\cdot)$ and $\psi_{\mathrm{cls}}(\cdot)$ are MLP-based decoders. The first head outputs reconstructed continuous edge attributes, while the second outputs logits over discrete semantic classes.

Unlike graph-level reconstruction or node-level anomaly detection, \SystemName{} directly models whether an individual communication edge behaves as expected under its structural context.

\subsection{Learning Objective}
\SystemName{} is trained exclusively on benign graphs. To force the model to infer edge behavior from graph context rather than memorizing raw edge attributes, we adopt a masked reconstruction objective. 
A standard autoencoder objective without masking would risk identity mapping, where the model simply copies the input features. 
By masking a subset of edges, we strictly force the model to genuinely rely on structural dependencies.

During training, we randomly sample a subset of edges $\mathcal{M} \subseteq E_t$ and mask their input edge attributes:

\begin{equation}
\tilde{e}_{ij} =
\begin{cases}
\mathbf{0}, & (v_i, v_j) \in \mathcal{M}, \\
e_{ij}, & \text{otherwise}.
\end{cases}
\end{equation}

Only masked edges contribute to the reconstruction loss. Let $y_{ij}^{\mathrm{reg}}$ and $y_{ij}^{\mathrm{cls}}$ denote the continuous and discrete ground-truth semantics of edge $e_{ij}$, respectively. The regression loss is defined as
\begin{equation}
\mathcal{L}_{\mathrm{reg}} =
\frac{1}{|\mathcal{M}|}
\sum_{(v_i, v_j) \in \mathcal{M}}
\ell_{\mathrm{reg}}\!\left(\hat{y}^{\mathrm{reg}}_{ij}, y^{\mathrm{reg}}_{ij}\right),
\end{equation}
where $\ell_{\mathrm{reg}}(\cdot,\cdot)$ is Smooth L1 loss. The classification loss is
\begin{equation}
\mathcal{L}_{\mathrm{cls}} =
\frac{1}{|\mathcal{M}|}
\sum_{(v_i, v_j) \in \mathcal{M}}
\ell_{\mathrm{CE}}\!\left(\hat{y}^{\mathrm{cls}}_{ij}, y^{\mathrm{cls}}_{ij}\right),
\end{equation}
where $\ell_{\mathrm{CE}}$ is cross-entropy loss over the discrete semantic categories.

The overall training objective is
\begin{equation}
\mathcal{L} =
\lambda_{\mathrm{reg}} \mathcal{L}_{\mathrm{reg}}
+
\lambda_{\mathrm{cls}} \mathcal{L}_{\mathrm{cls}},
\end{equation}
where $\lambda_{\mathrm{reg}}$ and $\lambda_{\mathrm{cls}}$ control the relative contributions of continuous and discrete semantics.

This objective encourages \SystemName{} to infer edge semantics from structural context rather than memorizing raw attributes. Under the benign-only setting, the resulting reconstruction function captures the regularities of normal communication behavior and provides the basis for one-class anomaly detection at inference time.

\subsection{Anomaly Scoring and Aggregation}

At inference time, \SystemName{} assigns an anomaly score to each communication edge by measuring how well the observed edge semantics agree with the model's structure-conditioned reconstruction. For the continuous part, we compute the reconstruction discrepancy as the Mean Absolute Error (MAE) across the regressed dimensions:
\begin{equation}
s^{\mathrm{reg}}_{ij} =
\frac{1}{d_e^{\mathrm{reg}}}
\sum_{k=1}^{d_e^{\mathrm{reg}}}
\left|
\hat{y}^{\mathrm{reg}}_{ij,k} - y^{\mathrm{reg}}_{ij,k}
\right|.
\end{equation}

For the discrete part, we use the confidence assigned to the true semantic label. Let $p_{ij}(y^{\mathrm{cls}}_{ij})$ denote the predicted probability of the correct discrete class. We define the discrete inconsistency score as
\begin{equation}
s^{\mathrm{cls}}_{ij} = 1 - p_{ij}(y^{\mathrm{cls}}_{ij}).
\end{equation}

Because raw reconstruction errors may vary across graphs, we calibrate both scores using robust statistics estimated from benign training edges. Let $\mathrm{med}(\cdot)$ and $\mathrm{MAD}(\cdot)$ denote the median and median absolute deviation, respectively. 

We then normalize the regression and classification discrepancies into robust $z$-scores. To improve numerical stability, we impose a lower bound $\tau_{\mathrm{MAD}}$ on the estimated MAD and clip the normalized scores to the range $[-\tau_{\mathrm{clip}}, \tau_{\mathrm{clip}}]$. The calibrated scores are defined as

\begin{equation}
z^{\mathrm{reg}}_{ij}
=
\mathrm{clip}\!\left(
\frac{s^{\mathrm{reg}}_{ij} - \mathrm{med}(s^{\mathrm{reg}})}
{\max(\mathrm{MAD}(s^{\mathrm{reg}}), \tau_{\mathrm{MAD}})},
-\tau_{\mathrm{clip}}, \tau_{\mathrm{clip}}
\right),
\end{equation}

\begin{equation}
z^{\mathrm{cls}}_{ij}
=
\mathrm{clip}\!\left(
\frac{s^{\mathrm{cls}}_{ij} - \mathrm{med}(s^{\mathrm{cls}})}
{\max(\mathrm{MAD}(s^{\mathrm{cls}}), \tau_{\mathrm{MAD}})},
-\tau_{\mathrm{clip}}, \tau_{\mathrm{clip}}
\right).
\end{equation}

The final edge anomaly score is defined as
\begin{equation}
s_{ij} = z^{\mathrm{reg}}_{ij} + \alpha z^{\mathrm{cls}}_{ij},
\end{equation}
where $\alpha$ controls the contribution of discrete semantic inconsistency. Large values of $s_{ij}$ indicate that the observed communication edge cannot be well explained by its structural context under the benign model.

To identify suspicious hosts, \SystemName{} aggregates edge-level anomaly scores to the node level. For each node $v_i$, we collect the scores of incident edges and compute
\begin{equation}
S(v_i) = \mathrm{Agg}\left(
\{ \lambda_{\mathrm{src}} s_{ij} \mid (v_i, v_j) \in E_t \}
\cup
\{ \lambda_{\mathrm{dst}} s_{ji} \mid (v_j, v_i) \in E_t \}
\right),
\end{equation}
where $\mathrm{Agg}(\cdot)$ denotes an aggregation operator over the weighted incident edge scores, and $\lambda_{\mathrm{src}} > \lambda_{\mathrm{dst}} \ge 0$ assign higher importance to outgoing suspicious communications than incoming ones. In this work, we consider four instantiations of $\mathrm{Agg}(\cdot)$: \textit{mean}, \textit{max}, \textit{q}-quantile, and \textit{top-k mean}. Here, \textit{q90} denotes the 90th percentile of the weighted incident scores, and \textit{top-k mean} averages the largest $\max(\lfloor 0.1m_i \rfloor, 1)$ scores for node $v_i$, where $m_i$ is the number of weighted incident edges collected for that node. This source-aware weighting reflects our C2-motivated assumption that compromised hosts often initiate suspicious communications, making source-side attribution more appropriate for host-level anomaly detection.

The same aggregation principle can be extended to graph- or window-level scoring when required. However, since \SystemName{} is currently evaluated as a host-prioritization system, the experiments in this paper focus on node-level detection rather than direct edge-level ranking metrics.

\section{Experiments}
We evaluate \SystemName{} in terms of overall detection performance and the contribution of its key design choices.

\subsection{Experimental Setup}
We evaluate \SystemName{} on two network intrusion datasets that are widely used in surveys of network-based NIDS benchmarks: CTU-13~\cite{ref_ctu13} and CICIDS2017~\cite{ref_cicids,ref_dataset_survey}. For both datasets, network traffic is converted into attributed communication graphs using the graph construction procedure described in the method section. Benign graph snapshots are ordered chronologically. The earliest 80\% form the benign training pool, while the remaining 20\% together with all malicious windows constitute the test set. Node-level evaluation follows the source-centric labeling rule adopted in our graph construction process.

The reported \SystemName{} checkpoints use the feature configuration summarized in Table~\ref{tab:feature_profile}, yielding 51-dimensional node features and 9-dimensional edge features. The model uses a two-layer GINE-based encoder with hidden size 128, reconstructs the three continuous edge statistics (communication count, bytes, and packets), and predicts the three-way destination-port bucket. Training uses dropout 0.2, edge masking ratio 0.2, AdamW with learning rate $10^{-3}$, a 10\% validation split drawn from the benign training graphs, and equal loss weights $\lambda_{\mathrm{reg}}=\lambda_{\mathrm{cls}}=1.0$. For score calibration, we use $\alpha=1.0$, a MAD floor of $\tau_{\mathrm{MAD}}=10^{-3}$, and clipped robust scores with $\tau_{\mathrm{clip}}=10.0$. Unless otherwise noted, we assign weight 1.0 to the source endpoint and 0.2 to the destination endpoint, and report the main comparison with q90 node aggregation as a shared default across datasets. We use q90 as the shared default because it preserves a consistent headline configuration across datasets while still delivering a strong cross-dataset balance: under the current manuscript results it retains the strongest PR-AUC on CTU-13 and the best ROC-AUC, PR-AUC, and TPR@1\%FPR on CICIDS2017, even though other aggregators can be preferable at different operating points. Here, q90 denotes the 90th percentile of the weighted incident scores; \textit{top-k mean} uses a ratio of 0.1 with a minimum of one edge. Section~4.3 analyzes how alternative aggregation operators shift the operating-point trade-offs. We report node-level ROC-AUC, PR-AUC, and True Positive Rate at strict False Positive Rate limits (TPR@1\%FPR and TPR@5\%FPR).

All baselines use the same chronological split and the same source-centric node labeling protocol. Isolation Forest~\cite{ref_if} is fit with 200 trees on node features from the constructed graph snapshots. The Autoencoder baseline uses the same node features after standardization and trains a two-layer MLP autoencoder (hidden size 128, latent size 32) for 30 epochs with Adam at learning rate $10^{-3}$ and batch size 1024; node scores are mean absolute reconstruction error. GraphSAGE+IF~\cite{ref_graphsage} uses the same graph snapshots and node features, trains a two-layer GraphSAGE encoder (hidden size 64, output size 64, dropout 0.2) for 20 epochs with self-supervised link prediction, and then fits an Isolation Forest with 200 trees on benign node embeddings. Kitsune~\cite{ref_kitsune} follows a KitNET-style ensemble with maximum autoencoder size 10 on standardized node features. Our Anomal-E baseline~\cite{ref_anomale} uses the same graph snapshots, trains an edge-aware DGI-style encoder with hidden and output size 64, dropout 0.2, and Adam at learning rate $10^{-3}$, fits an Isolation Forest with 200 trees on benign edge embeddings, and converts edge scores to node scores with max incident-edge aggregation. Unless otherwise noted, all reported results use a single random seed.

\subsection{Overall Performance}
\begin{table*}[t]
\centering
\caption{Node-level anomaly detection performance on CTU-13 and CICIDS2017. Bold denotes the best performance.}
\label{tab:main_results}
\resizebox{\textwidth}{!}{
\begin{tabular}{l|cccc|cccc}
\toprule
\multirow{2}{*}{Method} 
& \multicolumn{4}{c|}{CTU-13} 
& \multicolumn{4}{c}{CICIDS2017} \\
\cline{2-9}
& ROC-AUC $\uparrow$ & PR-AUC $\uparrow$ & TPR@1\%FPR $\uparrow$ & TPR@5\%FPR $\uparrow$
& ROC-AUC $\uparrow$ & PR-AUC $\uparrow$ & TPR@1\%FPR $\uparrow$ & TPR@5\%FPR $\uparrow$ \\
\midrule
Isolation Forest 
& 0.8750 & 0.0473 & 0.0216 & 0.0550 
& 0.8890 & 0.1511 & 0.1471 & 0.5252 \\

Autoencoder 
& 0.9443 & 0.1126 & 0.0234 & 0.6419 
& 0.9285 & 0.2983 & 0.3869 & 0.5960 \\

GraphSAGE + IF 
& 0.7636 & 0.0254 & 0.0017 & 0.0300 
& 0.5350 & 0.0141 & 0.0150 & 0.0974 \\

Kitsune 
& 0.9508 & \textbf{0.1713} & 0.0147 & 0.7566 
& 0.7493 & 0.1427 & 0.1519 & 0.4802 \\

Anomal-E 
& 0.7661 & 0.0300 & \textbf{0.0403} & 0.1032 
& 0.8728 & 0.0820 & 0.1315 & 0.3951 \\

\midrule
\SystemName{} (Ours)
& \textbf{0.9804} & 0.1618 & 0.0302 & \textbf{0.9647}
& \textbf{0.9753} & \textbf{0.3175} & \textbf{0.4463} & \textbf{0.8569} \\

\bottomrule
\end{tabular}
}
\end{table*}

Table~\ref{tab:main_results} reports the main node-level detection results of \SystemName{} and all baselines on CTU-13 and CICIDS2017. For \SystemName{}, the reported results use the shared default setting described above: source-weighted edge-to-node attribution with $\alpha=1.0$ and q90 node aggregation. We use this shared configuration to keep the headline comparison fixed across datasets, while Table~\ref{tab:agg_ablation} later compares alternative aggregation operators and their dataset-specific trade-offs.

On CTU-13, \SystemName{} achieves the best ROC-AUC (0.9804) and TPR@5\%FPR (0.9647) among the evaluated methods. It also attains competitive PR-AUC (0.1618), second only to Kitsune (0.1713), but it does not achieve the best TPR@1\%FPR, where Anomal-E remains slightly higher (0.0403 versus 0.0302). This pattern suggests that \SystemName{} is strongest as a host-ranking method around the 5\% FPR operating point, while its advantage is less uniform at the strictest threshold.

On CICIDS2017, \SystemName{} achieves the best performance among the evaluated methods on all reported metrics, including ROC-AUC (0.9753), PR-AUC (0.3175), TPR@1\%FPR (0.4463), and TPR@5\%FPR (0.8569). The margin at the strict false-positive operating points is substantial, indicating that \SystemName{} can rank suspicious hosts more effectively than both classical feature-based methods and the compared graph-based baselines when the alert budget is limited.

Taken together, these results indicate that modeling structure-conditioned edge semantics can improve benign-only host ranking under strict false-positive constraints, although the gains depend on the dataset and the operating point.

\subsection{Ablation Studies}

\paragraph{Effect of MAD Calibration}

Raw reconstruction errors can vary substantially across different graph neighborhoods, making raw scores unreliable for global ranking. Table~\ref{tab:mad_ablation} compares raw reconstruction scores against MAD-calibrated scores on both datasets. On CTU-13, calibration substantially improves ROC-AUC, PR-AUC, and TPR@5\%FPR, although TPR@1\%FPR decreases slightly. On CICIDS2017, calibration consistently improves all reported metrics. Overall, these results indicate that robust MAD calibration is important for translating edge-level reconstruction discrepancies into stable node-level anomaly rankings under strict false positive constraints.

\begin{table*}[t]
\centering
\caption{Effect of robust MAD calibration on node-level detection performance. For each dataset, ``raw'' denotes uncalibrated reconstruction scores and ``calibrated'' denotes the final calibrated setting used by \SystemName{}. Bold denotes the better result within each dataset and metric.}
\label{tab:mad_ablation}
\resizebox{\textwidth}{!}{
\begin{tabular}{l|l|cccc}
\toprule
Dataset & Setting & ROC-AUC $\uparrow$ & PR-AUC $\uparrow$ & TPR@1\%FPR $\uparrow$ & TPR@5\%FPR $\uparrow$ \\
\midrule
\multirow{2}{*}{CTU-13}
& raw        & 0.8855 & 0.0815 & \textbf{0.0428} & 0.4506 \\
& calibrated & \textbf{0.9804} & \textbf{0.1618} & 0.0302 & \textbf{0.9647} \\
\midrule
\multirow{2}{*}{CICIDS2017}
& raw        & 0.9115 & 0.2659 & 0.3379 & 0.5177 \\
& calibrated & \textbf{0.9753} & \textbf{0.3175} & \textbf{0.4463} & \textbf{0.8569} \\
\bottomrule
\end{tabular}
}
\end{table*}

\paragraph{Effect of Aggregation Strategy}

Because \SystemName{} evaluates individual communication edges, edge-level scores must be aggregated to identify compromised hosts. Table~\ref{tab:agg_ablation} compares four aggregation operators introduced in Section~3.5: \textit{mean}, \textit{q90}, \textit{max}, and \textit{top-k mean}. To keep the headline comparison fixed, the q90 row matches the shared-default \SystemName{} result reported in Table~\ref{tab:main_results}, while the remaining rows report additional runs with alternative node aggregation operators under the same source-weighted setting. On CTU-13, the aggregation choice changes which operating point is favored. The fixed q90 default yields the strongest PR-AUC, \textit{top-k mean} yields the highest TPR@1\%FPR, and both \textit{max} and \textit{top-k mean} reach the highest TPR@5\%FPR, with \textit{max} also producing the best ROC-AUC. On CICIDS2017, \textit{q90} remains best on ROC-AUC, PR-AUC, and TPR@1\%FPR, while \textit{max} provides a small TPR@5\%FPR gain. These results reinforce that the optimal aggregation strategy is dataset- and operating-point-dependent.

\begin{table*}[t]
\centering
\caption{Effect of edge-to-node aggregation strategy on node-level detection performance. The q90 row matches the shared-default \SystemName{} result reported in Table~\ref{tab:main_results}; the remaining rows use alternative node aggregation operators under the same source-weighted setting. Bold denotes the best performance for each dataset and metric.}
\label{tab:agg_ablation}
\resizebox{\textwidth}{!}{
\begin{tabular}{l|cccc|cccc}
\toprule
\multirow{2}{*}{Aggregation}
& \multicolumn{4}{c|}{CTU-13}
& \multicolumn{4}{c}{CICIDS2017} \\
\cline{2-9}
& ROC-AUC $\uparrow$ & PR-AUC $\uparrow$ & TPR@1\%FPR $\uparrow$ & TPR@5\%FPR $\uparrow$
& ROC-AUC $\uparrow$ & PR-AUC $\uparrow$ & TPR@1\%FPR $\uparrow$ & TPR@5\%FPR $\uparrow$ \\
\midrule
mean
& 0.9691 & 0.1068 & 0.0251 & 0.9506
& 0.9664 & 0.2531 & 0.3917 & 0.8350 \\

q90
& 0.9804 & \textbf{0.1618} & 0.0302 & 0.9647
& \textbf{0.9753} & \textbf{0.3175} & \textbf{0.4463} & 0.8569 \\

max
& \textbf{0.9860} & 0.1166 & 0.0283 & \textbf{0.9929}
& 0.9731 & 0.1892 & 0.3052 & \textbf{0.8668} \\

top-k mean
& 0.9854 & 0.0159 & \textbf{0.1294} & \textbf{0.9929}
& 0.9744 & 0.2181 & 0.3280 & 0.8588 \\
\bottomrule
\end{tabular}
}
\end{table*}

\section{Discussion}

The experiments support three main observations. First, \SystemName{} is most compelling as a host-ranking method under constrained false-positive budgets rather than a universally dominant detector. On CTU-13 it achieves the best ROC-AUC and TPR@5\%FPR and also delivers competitive PR-AUC, but it does not lead at TPR@1\%FPR. On CICIDS2017, by contrast, \SystemName{} leads on all reported metrics. Second, MAD calibration materially changes the ranking behavior of reconstruction scores. The gain is especially large on CTU-13, where calibration sharply improves ROC-AUC, PR-AUC, and TPR@5\%FPR while slightly reducing TPR@1\%FPR. Third, edge-to-node aggregation is not a cosmetic post-processing step. On CTU-13, different operators favor different alert budgets: q90 preserves the strongest PR-AUC, top-k mean achieves the highest TPR@1\%FPR, and max and top-k mean achieve the highest TPR@5\%FPR, with max also producing the strongest ROC-AUC. On CICIDS2017, q90 remains best on ROC-AUC, PR-AUC, and TPR@1\%FPR, while max gives a small TPR@5\%FPR gain. These patterns suggest that \SystemName{}'s benefits depend on how edge-level inconsistencies are normalized and mapped to host-level alerts. 
In our framework, we adopted source-centric node labeling and attribution. 
This choice is justified by our motivating threat model, where compromised hosts actively initiate C2 communications. 
In contrast, while symmetric attribution is common in general graph anomaly detection, it has drawbacks such as diluting the anomaly score of the actual attacker, and was therefore not used as our default.

\textbf{Limitations.} We acknowledge several constraints in our current evaluation. 
First, our findings rely on single-seed offline benchmarks and historical traffic traces rather than live, real-time deployments. 
Second, although \SystemName{} is motivated as an edge-centric model, the evidence reported here is host-level: we do not yet include direct edge-level ranking metrics or analyst-facing case studies for suspicious communications. 
Third, our host labels and aggregation rule are explicitly source-centric. We mark only the initiator of a malicious flow as anomalous and use source-weighted edge aggregation, which reflects a C2-motivated threat model but may not transfer unchanged to settings where destination-side evidence or symmetric attribution is equally important. 
However, a key consideration is that \SystemName{} currently processes static time windows sequentially. 
This reset of state between windows potentially limits the detection of ultra-slow, cross-window beaconing. 
Future research could be directed towards online updating and streaming GNN architectures to address this.
Finally, long-term deployment in dynamic enterprise environments will inevitably encounter concept drift, where the benign baseline shifts over time. 
Addressing such shifts would require periodic retraining or incremental recalibration of the MAD statistics. 
Future work should, therefore, consider online updating, streaming inference, edge-level validation, and evaluation under long-term non-stationary traffic.

\section{Related Work}

\noindent\textbf{Feature-Based Detection.}
Traditional intrusion detection has long relied on supervised learning over flow-level or packet-level features, using models such as Random Forests, Support Vector Machines, and Multi-Layer Perceptrons trained on labeled traffic traces~\cite{ref_supervised_limits,ref_ids_survey,ref_buczak}. These methods are often effective for known attack signatures and benchmark settings, but their performance can degrade when attack behaviors evolve or when labeled malicious traffic is scarce. This limitation is especially relevant when suspicious communications are intentionally designed to resemble benign traffic and may therefore be underrepresented in historical attack datasets.

\noindent\textbf{Reconstruction-Based Detection.}
To reduce reliance on attack labels, a large body of work has explored unsupervised anomaly detection for network traffic. Classical approaches such as Isolation Forest and related outlier detection methods aim to identify deviations from dominant traffic patterns~\cite{ref_anomaly_survey,ref_if}. Reconstruction-based models further learn a compact representation of normal traffic and detect anomalies through reconstruction error. Autoencoder-based methods follow this paradigm, while systems such as Kitsune~\cite{ref_kitsune} employ an ensemble of autoencoders to model packet- or flow-level statistics in an online setting. Studies of encrypted traffic further show that TLS metadata, contextual side information, and careful feature engineering can provide strong detection signal without decryption~\cite{ref_tls_context,ref_tls_nonstationary,ref_tls_usage}. Although these methods are attractive in benign-only scenarios, they typically operate on independent traffic records and do not explicitly capture relational communication structure among hosts.

\noindent\textbf{Graph-Based Detection.}
Security researchers exploited correlation and structure in botnet traffic well before the recent wave of graph neural intrusion detection. BotSniffer used spatial-temporal correlation to detect botnet C2 channels~\cite{ref_botsniffer}, BotMiner correlated communication and malicious activity clusters to recover bots independent of protocol and structure~\cite{ref_botminer}, and BotGrep localized P2P bots from structured overlay communication patterns~\cite{ref_botgrep}. More recent work has introduced graph-based representations to better model dependencies among communicating entities. Early graph-based intrusion detection approaches apply GraphSAGE-, GCN-, or embedding-based models to host interaction graphs, learning node- or graph-level representations from neighborhood connectivity~\cite{ref_graphsage,ref_graph_based}. Beyond security-specific systems, generic graph anomaly detectors such as DOMINANT~\cite{ref_dominant}, CoLA~\cite{ref_cola}, and ANEMONE~\cite{ref_anemone} show that reconstruction- and contrastive-learning objectives can identify anomalous nodes or substructures in attributed graphs. Other studies extend graph learning to encrypted or malware-oriented settings by modeling traffic as spatial-temporal graphs and applying supervised graph learning to identify malicious activity despite protocol obfuscation~\cite{ref_stgraph}. Temporal and streaming graph methods further emphasize continuous monitoring and low-latency anomaly tracking in dynamic graph streams~\cite{ref_edgetorrent}. More recently, self-supervised graph methods such as Anomal-E~\cite{ref_anomale} have shown that edge-aware graph representation learning can improve intrusion detection without requiring attack labels.

\SystemName{} differs from these lines of work in three concrete ways. Compared with Kitsune-style reconstruction, it conditions each communication on graph context rather than treating traffic records independently. Compared with GraphSAGE-style host embeddings, it treats the communication edge rather than the host as the primary detection object. Compared with Anomal-E, which learns edge-aware embeddings for downstream anomaly detection, \SystemName{} reconstructs explicit edge semantics and calibrates the resulting discrepancies with MAD before host-level ranking. The method therefore targets a narrower question: whether a specific communication is compatible with the structural role implied by its local neighborhood.

\section{Conclusion}

Benign-only detection from flow logs remains challenging when suspicious communications are designed to resemble normal traffic. We presented \SystemName{}, a graph-based framework that models network activity as attributed communication graphs and detects anomalies through structure-conditioned edge semantic reconstruction. 
By reconstructing both continuous and discrete edge semantics from structural context, \SystemName{} identifies communications whose observed behavior does not match their local graph role.
To the best of our knowledge, \SystemName{} is the first framework to formulate benign-only flow-log intrusion detection explicitly through structure-conditioned edge semantic reconstruction.
Our evaluations on CTU-13 and CICIDS2017 show that \SystemName{} is most convincing under constrained false-positive budgets, with especially strong performance on CICIDS2017 and a favorable ranking profile on CTU-13, including the best ROC-AUC and TPR@5\%FPR together with competitive PR-AUC. The results do not establish a universal advantage across datasets or operating points, but they do show that edge-level semantic reconstruction, robust calibration, and careful aggregation can materially improve host ranking in benign-only flow-log intrusion detection.
Future work will investigate online updating, temporal graph modeling, and adaptive calibration under concept drift to extend \SystemName{} toward continuous streaming environments.

%
%

\end{document}